\documentclass{article} % For LaTeX2e
\usepackage{nips13submit_e,times}
\usepackage{hyperref}
\usepackage{url}
\usepackage{natbib}

\title{Evolution of the user's content: An Overview of the state of the art}

\author{
Djallel Bouneffouf \\
Department of Computer Science\\
Telecom SudParis\\
\texttt{Djallel.bouneffouf@it-sudparis.eu}
}
\begin{document}

\maketitle

\begin{abstract}
The  evolution of the user's content still remains a problem for an accurate recommendation.This is why the current research aims to design \textit{Recommender Systems (RS)} able to continually adapt information that matches the user's interests. This paper aims to explain this problematic point in outlining the proposals that have been made in research with their advantages and disadvantages.    
\end{abstract}

\section{Introduction}

The  evolution of the user's content still remains a problem for an accurate recommendation. This section aims to explain this problematic point in outlining the proposals that have been made in research with their advantages and disadvantages.
\\
The evolution of the user's content is reflected in the difficulty of generating recommendations for the user where old documents may expire and new documents may emerge continuously. This context lets the user's profile in a permanent cold start, also called latancy \cite{Schein2002}. 
\\
The cold start problem occurs, in fact, when a new document is integrated into the system and the relation between the user's interests and the document is not yet available. Therefore, the new document will not be involved in the recommendations. The problem of latency has more impact on systems that incorporate new documents regularly, such as systems recommending news articles \cite{13} \cite{Burke}.
\\
To overcome this latency problem, two approaches are proposed, the first one uses CBF and the second one uses an exploration / exploitation strategy. These two approaches are described in what follows.

\section{Content-based Filtering}

Content-based filtering try to solve cold start by technique based on content similarities \cite{Pazzani}. When a new document is introduced, the technique based on the content evaluates the similarity of the new document with the available documents in the user's profiles to make recommendation. However, the use of the technique based on the content leads to a lack of diversity in recommendations, which hinders the performance of the recommendation system.
\\
A new filtering technique (based on content) exploiting ontologies has also been suggested as a solution to the problem of latency. This technique has been especially used by \cite{Burke} to recommends restaurants and by \cite{ontologiesprofile} to recommend scientific papers. These systems use location ontologies, in order to classify and categorize documents and generate the users profiles. Therefore, the limitation of this technique is the need to pre-construct the domain knowledge ontology.
\\
In \cite{Burke}, the authors propose a system where the user's interests at different moments should have different weights. To catch a user's short-term interests without losing the long-term interests, the authors propose a novel CF approach to consider the user's contents evolution. There are four major phases in the proposed approach: document clustering, user interest derivation, neighborhood formation, and recommendation generation.
\\ 
The phase of document clustering groups documents with similar contents by adopting a specific cluster algorithm. 
\\
The phase of user's interests derivation calculates the weight of each document based on the given user's preferences on each document as well as the corresponding timestamps. Then, the interest vector of each user is constructed to represent the user's current short-term interests .
\\
After the user interest derivation, the phase of neighborhood formation identifies the neighbors of the active user based on the similarity of their  interest vectors. 
\\
Finally, the recommendation generation phase estimates the interests for the active user, to take into account in the document selection, based on the opinions of his/her neighbors. 
\\
Despite their interest, these approaches have some limitations including lack diversity in the recommendations or deterioration of the quality of recommendations due to the use of social information.

\section{Exploration / Exploitation Trade-off}
Few works found in the literature \cite{13, 21} solve the problem of content evolution by addressing it as a need for balancing exploration and exploitation studied in the “multi-armed bandit problem”. 
\\
A bandit algorithm B exploits its past experience to select documents (arms) that appear more frequently. Besides, these seemingly optimal documents may in fact be suboptimal, because of the imprecision in B's knowledge. In order to avoid this undesired situation, B has to explore documents by choosing seemingly suboptimal documents so as to gather more information about them.
Exploration can decrease short-term user's satisfaction when some suboptimal documents may be chosen. However, obtaining information about the documents' average rewards can refine B's estimate of the documents' rewards and in turn increases the long-term user's satisfaction. 
\\
Clearly, neither a purely exploring nor a purely exploiting algorithm works well, and then a good trade-off is needed.
\\
The exploration / exploitation approach is used in \cite{13} for ads recommendation and in \cite{21} for news recommendation. 

\section{Conclusion and Discussion}
To discuss the two approaches for following the user's content evolution, we present, in table ~\ref{tab:FUIE}, their advantages and disadvantages. 

\begin{table} [h]
\caption{ Approaches for following the user's content evolution}
\label{tab:FUIE}       % Give a unique label
\begin{tabular}{|p{3cm}|p{5cm}|p{5cm}|p{3cm}|}
\hline
\bf Approach         & \bf Advantage & \bf Disadvantage 
 \\
\hline Content Filtering \cite{Burke} \cite{ontologiesprofile} & Recommendations similar to the user's history  & The lack of diversity in recommendations.
 \\
\hline Exploration / Exploitation Trade-off \cite{21} \cite{13} & Recommending diversified information to the user  & The risk to upset the user with randomly choose documents   \\
\hline 
\end{tabular}
\end{table}

From Table~\ref{tab:FUIE}, we can observe that the techniques based on content and CF have the advantage of exploiting to user's history in the recommendation. However, these techniques have a lack of diversity of content in recommendations.
\\
In the other case, Exploration / Exploitation Trade-off approaches have allowed recommending varied and diverse documents to the user, but the risk is to upset the user with these randomly chosen documents.
\\
We can conclude that a trade-off has to be done among the random exploration, the content and CF.
\\
Moreover, in order to be well accepted by the user, random recommendations should be managed regarding the user's context. 

\bibliographystyle{named}
\bibliography{profile}

\begin{thebibliography}{}

\bibitem[\protect\citeauthoryear{Abowd \bgroup \em et al.\egroup }{1999}]{Dey}
Gregory~D. Abowd, Anind~K. Dey, Peter~J. Brown, Nigel Davies, Mark Smith, and
  Pete Steggles.
\newblock Towards a better understanding of context and context-awareness.
\newblock In {\em Proceedings of the 1st international symposium on Handheld
  and Ubiquitous Computing}, HUC '99, pages 304--307, London, UK, UK, 1999.
  Springer-Verlag.

\bibitem[\protect\citeauthoryear{Adomavicius and Tuzhilin}{2005}]{adomavicus}
Gediminas Adomavicius and Alexander Tuzhilin.
\newblock Toward the next generation of recommender systems: A survey of the
  state-of-the-art and possible extensions.
\newblock {\em IEEE Trans. on Knowl. and Data Eng.}, 17(6):734--749, June 2005.

\bibitem[\protect\citeauthoryear{Bogers and van~den Bosch}{2009}]{hybridCF}
Toine Bogers and Antal van~den Bosch.
\newblock Collaborative and content-based filtering for item recommendation on
  social bookmarking websites, 2009.

\bibitem[\protect\citeauthoryear{Burke}{2002}]{Burke}
Robin Burke.
\newblock Hybrid recommender systems: Survey and experiments.
\newblock {\em User Modeling and User-Adapted Interaction}, 12(4):331--370,
  November 2002.

\bibitem[\protect\citeauthoryear{Grudin}{2001}]{Grudin}
Jonathan Grudin.
\newblock Partitioning digital worlds: focal and peripheral awareness in
  multiple monitor use.
\newblock In {\em CHI}, pages 458--465, 2001.

\bibitem[\protect\citeauthoryear{Li \bgroup \em et al.\egroup }{2010a}]{13}
Lihong Li, Wei Chu, John Langford, and Robert~E. Schapire.
\newblock A contextual-bandit approach to personalized news article
  recommendation.
\newblock In {\em Proceedings of the 19th international conference on World
  wide web}, WWW '10, pages 661--670, USA, 2010. ACM.

\bibitem[\protect\citeauthoryear{Li \bgroup \em et al.\egroup }{2010b}]{21}
Wei Li, Xuerui Wang, Ruofei Zhang, and Ying Cui.
\newblock Exploitation and exploration in a performance based contextual
  advertising system.
\newblock In {\em Proceedings of the 16th ACM SIGKDD international conference
  on Knowledge discovery and data mining}, KDD '10, pages 27--36, USA, 2010.
  ACM.

\bibitem[\protect\citeauthoryear{Middleton \bgroup \em et al.\egroup
  }{2004}]{ontologiesprofile}
Stuart~E. Middleton, Nigel~R. Shadbolt, and David~C. De~Roure.
\newblock Ontological user profiling in recommender systems.
\newblock {\em ACM Trans. Inf. Syst.}, 22(1):54--88, January 2004.

\bibitem[\protect\citeauthoryear{Pazzani and Billsus}{2007}]{Pazzani}
Michael~J. Pazzani and Daniel Billsus.
\newblock Content-based recommendation systems.
\newblock In {\em The Adaptive Web}, pages 325--341, 2007.

\bibitem[\protect\citeauthoryear{Sarwar \bgroup \em et al.\egroup }{2001}]{CF}
Badrul Sarwar, George Karypis, Joseph Konstan, and John Riedl.
\newblock Item-based collaborative filtering recommendation algorithms.
\newblock In {\em Proceedings of the 10th international conference on World
  Wide Web}, WWW '01, pages 285--295, New York, NY, USA, 2001. ACM.

\bibitem[\protect\citeauthoryear{Schein \bgroup \em et al.\egroup
  }{2002}]{Schein2002}
Andrew~I. Schein, Alexandrin Popescul, Lyle~H. Ungar, and David~M. Pennock.
\newblock Methods and metrics for cold-start recommendations.
\newblock In {\em Proceedings of the 25th annual international ACM SIGIR
  conference on Research and development in information retrieval}, SIGIR '02,
  pages 253--260, New York, NY, USA, 2002. ACM.

\bibitem[\protect\citeauthoryear{Shani \bgroup \em et al.\egroup
  }{2007}]{Profile}
Guy Shani, Lior Rokach, Amnon Meisles, Lihi Naamani, Nischal Piratla, and David
  Ben-shimon.
\newblock Establishing user profiles in the mediascout recommender system,
  2007.

\bibitem[\protect\citeauthoryear{van Meteren and van
  Someren}{}]{contentfiltering}
Robin van Meteren and Maarten van Someren.
\newblock {Using Content-Based Filtering for Recommendation}.

\end{thebibliography}

\end{document}